\documentclass[12pt,  onecolumn]{article}
\usepackage{amsmath}
\usepackage{amssymb}
\usepackage{graphicx}
\usepackage{caption2}
\usepackage{amsfonts}
\usepackage{cite}

\newcommand{\be}{\begin{equation}}
\newcommand{\ee}{\end{equation}}
\newcommand{\bea}{\begin{eqnarray}}
\newcommand{\eea}{\end{eqnarray}}
\newcommand{\bef}{\begin{figure}}
\newcommand{\ef}{\end{figure}}
\newcommand{\bt}{\begin{tabular}}
\newcommand{\et}{\end{tabular}}
\newcommand{\bno}{\begin{enumerate}}
\newcommand{\eno}{\end{enumerate}}

\def\3{\ss}

\catcode`\"=\active
\def"{\accent'177}

\begin{document}

\begin{center}

{\bf \Large  A comment on the arguments about the reliability  and convergence of chaotic simulations}

\vspace{0.3cm}

Shijun  Liao $^{a,b}$ 

\vspace{0.25cm}

$^a$   Key Laboratory of Education-Ministry in Scientific Computing,   Shanghai 200240, China

 \vspace{0.25cm}
 $^b$  School of Naval Architecture, Ocean and Civil Engineering\\
 Shanghai Jiao Tong University, Shanghai 200240, China\\

\end{center}

{\bf Abstract}  {\em \small Yao  and  Hughes  commented  (Tellus-A, 60:  803 -- 805, 2008) that ``all chaotic responses are simply numerical noise and have nothing to do with the solutions of differential equations''.  However, using 1200 CPUs of the National Supercomputer TH-A1 and a parallel integral algorithm of the so-called ``Clean Numerical Simulation'' (CNS) based on the 3500th-order Taylor expansion and data in 4180-digit multiple precision, one can gain reliable, convergent chaotic solution of Lorenz equation in a rather long interval [0,10000].  This supports Lorenz's  optimistic viewpoint (Tellus-A, 60: 806 -- 807, 2008): ``numerical approximations can converge to a
chaotic true solution throughout any finite range of time''.} 

\setlength{\parindent}{0.75cm} 

Using a digit computer, Lorenz  \cite{Lorenz1963} found the famous ``butterfly-effect'' of  ``deterministic non-periodic'' solution of three-coupled ordinary differential equations, called  today  the Lorenz equation: the so-called chaotic solutions are rather sensitive to initial conditions.  This work was a milestone in the field of nonlinear dynamics.   In 1999,  Tucker \cite{Tucker1999, Tucker2002} further proved that the Lorenz equation supports a strange attractor.  Tucker's  work  is  a great  breakthrough  that  provides a positive answer to the Smale's 14th Problem \cite{Smale1998}.      

However, it was {\em practically} difficult to gain a reliable numerical simulation of   chaotic dynamic systems in any a given finite range of time,    
mainly because Lorenz \cite{Lorenz1989, Lorenz2006}  further found  that chaotic solutions are sensitive not only to initial conditions but also to numerical algorithms:   different numerical algorithms and different time-steps may lead to completely different numerical simulations of chaos.   For example, chaotic numerical simulations of Lorenz equation given by different traditional procedures were often repeatable only in a interval of time less than 30 Lorenz time unit (LTU).    So, ``computed'' dynamic behaviors observed for a finite time step in some non-linear discrete-time difference equations sometimes might have nothing to do with the ``exact'' solution of the original continuous-time differential equations at all, as confirmed  by some other researchers  \cite{Li2001, Teixeira2007}.  This numerical phenomenon lead to intense arguments \cite {Yao2008, Lorenz2008} about reliability of numerical simulations of chaotic dynamic systems.   Especially,  Yao and Hughes  \cite {Yao2008}  believed that ``all chaotic responses are simply numerical noise and have nothing to do with the solutions of differential equations''.    On the other side, using double precision data and a few examples based on the 15th-order Taylor-series procedure \cite{Corliss1982, Barrio2005} with decreasing time-step,   Lorenz \cite{Lorenz2008} was optimistic and believed that ``numerical approximations can converge to a chaotic true solution throughout any finite range of time, although, if the range is large, confirming the convergence can be utterly impractical.''    

Currently,    using the arbitrary-order Taylor series method (TSM) \cite{Corliss1982, Barrio2005} and data in arbitrary-precision,   Liao  \cite{Liao2009, Liao2013, Liao2014}  proposed the so-called ``Clean Numerical Simulation'' (CNS) to gain convergent, reliable chaotic results in a long but finite interval of time $[0,T]$.    Let $s(M,N)$ denote a numerical simulation of a nonlinear dynamic system given by the CNS,  where $M$  denotes the order of TSM and  $N$ the number of  digit  precision of data, respectively.   Here, the ``convergence'' means that, for a given interval $[0,T]$ with a properly chosen time-step $\Delta t$,  there exist a critical order $M^*$ of the TSM and a critical integer $N^*$ for digit precision such that all numerical simulations $s(M,N)$  given by the CNS are the same, i.e. with negligible differences, as long as $M>M^*$ and $N>N^{*}$.  This is mainly because truncation error and round-off error can be reduced  to  a  required,  rather small level as  long as  $M$ for the order of TSM  and $N$ for the digit precision are large enough.   Unlike traditional numerical algorithms for chaotic systems, the CNS searches for the critical order $M^*$ of the TSM and the critical $N^*$ of digit-precision for a given interval $[0,T]$ and a chosen time step $\Delta t$.         

In 2009, using the CNS with the 400th-order Taylor series method (TSM)  and data in 800-digit precision (by means of the computer algebra Mathematica with the time step $\Delta t=0.01$),  Liao \cite{Liao2009} gained, for the first time, a reliable chaotic solution
 of Lorenz equation in a long interval [0,1000] of time.     As  reported  by  Liao  \cite{Liao2009},  for a given interval $[0,T]$,  one can gain reliable, convergent chaotic simulations of Lorenz equation by means of the CNS with the $M$th-order TMS and data in $N$-digit precision, where $M>M^* \approx T/3$ and $N > N^* \approx 2 T/5$  in the case of $\Delta t = 0.01$.   Using the multiple precision (MP) library \cite{MP1990} and parallel computation, Wang et al.  \cite{Wang2011}  confirmed the reliability of Liao's chaotic solution in [0,1000] and gained a reliable chaotic result of Lorenz equation in [0,2500] by means of the CNS based on the 1000th-order TSM and data in 2100-digit precision (with $\Delta t=0.01$).   Currently,  Using 1200 CPUs of the National Supercomputer TH-A1 and a parallel integral algorithm of the CNS  based on the 3500th-order Taylor expansion and data in the 4180-digit multiple precision,  Liao and Wang  \cite{Liao-Wang2014}  obtained  a reliable chaotic solution of Lorenz equation in a rather long interval $0 \leq  t \leq  10000$:  its reliability and convergence was  further  confirmed  by means of the CNS using the 3600th-order TSM and the data in 4515-digit multiple precision.    To the best of my knowledge,  such a  convergent, reliable chaotic solution of Lorenz equation in such a long interval has never been reported, which provides us a numerical benchmark of reliable  chaotic  solution of dynamic systems.  
 
Currently,  Kehlet and Logg  \cite{Logg2013}  gained  a  reliable  chaotic  solution of  Lorenz equation  on the time interval [0, 1000] using the 200-order {\em finite element method} and data in 400-digits precision.   Its reliability was confirmed by means of the CNS with the 400th-order TSM and data in 800-digit precision.  Note that the CNS is a kind of {\em finite difference method}.   Therefore,  reliable, convergent chaotic results of Lorenz equation in a long interval [0,1000] can be indeed  obtained  by the  two completely  {\em different} numerical approaches!   All  of these support Lorenz's optimistic viewpoints:``numerical approximations can converge to a
chaotic true solution throughout any finite range of time''.        

Thus,  the ``Smale's 14th Problem''  has a prefect answer:  the  strange attractor of Lorenz equation not only exists, but also can be calculated accurately!           
    

\end{document}